\documentstyle[aps,eqsecnum,twocolumn,epsfig]{revtex}
\newcommand {\pr}{\partial}

\newcommand {\bg}{\bar \gamma}
\newcommand {\bp}{\bar \psi}

\begin{document}
\title{On the role of $\Lambda$-term in the evolution of Bianchi-I 
cosmological model with nonlinear spinor field} 
\author{Bijan Saha\\ 
Laboratory of Information Technologies\\ 
Joint Institute for Nuclear Research, Dubna\\ 
141980 Dubna, Moscow region, Russia\\ 
e-mail:  saha@thsun1.jinr.ru, bijan@cv.jinr.ru}
\author{G.N. Shikin\\
Department of Theoretical Physics\\
Russian Peoples' Friendship University\\
6, Miklukho maklay Street\\
117198 Moscow, Russia}
\maketitle 

\begin{abstract}
   Self-consistent solutions to nonlinear spinor field  equations 
in General Relativity are studied for the case of Bianchi  type-I 
space-time. It has been shown that introduction of $\Lambda$-term
in the Lagrangian generates oscillations of the Bianchi type-I model.
\end{abstract}

\section{Introduction}

The quantum field theory in curved space-time has been a matter of
great interest in recent years because of its applications to 
cosmology and astrophysics. The evidence of existence of strong
gravitational fields in our Universe led to the study of the quantum 
effects of material fields in external classical gravitational field.
After the appearance of Parker's paper on scalar fields~\cite{parker1} and 
spin-$\frac{1}{2}$ fields,~\cite{parker2} several authors have studied this 
subject. Although the Universe seems homogenous and isotropic at present,
there are no observational data guarantying the isotropy in the era prior
to the recombination. In fact, there are theoretical arguments that sustain 
the existence of an anisotropic phase that approaches an isotropic one
~\cite{misner}. Interest in studying Klein-Gordon and Dirac equations in 
anisotropic models has increased since Hu and Parker~\cite{hu1} have 
shown that the creation of scalar particles in anisotropic backgrounds can
dissipate the anisotropy as the Universe expands. 

A Bianchi type-I (B-I) Universe, being the straightforward generalization 
of the flat Robertson-Walker (RW) Universe, is one of the simplest models  
of an anisotropic Universe that describes a homogenous and spatially flat
Universe. Unlike the RW Universe which has the same 
scale factor for each of the three spatial directions, a B-I Universe
has a different scale factor in each direction, thereby introducing an
anisotropy to the system. It moreover has the agreeable property that
near the singularity it behaves like a Kasner Universe, even in the 
presence of matter, and consequently falls within the general analysis
of the singularity given by Belinskii et al~\cite{belinskii}. 
Also in a Universe filled with matter for $p\,=\,\gamma\,\varepsilon, \quad 
\gamma < 1$, it has been shown that any initial anisotropy in a B-I
Universe quickly dies away and a B-I Universe eventually evolves
into a RW Universe~\cite{jacobs}. Since the present-day Universe is 
surprisingly isotropic, this feature of the B-I Universe makes it a prime 
candidate for studying the possible effects of an anisotropy in the early 
Universe on present-day observations. In light of the importance of mentioned 
above, several authors have studied linear spinor field 
equations~\cite{chimento,castagnino} and the behavior of gravitational 
waves (GW's)~\cite{hu2,miedema,cho}    
in a B-I Universe. Nonlinear spinor field (NLSF) in external cosmological
gravitational field was first studied by G.N. Shikin in 1991~\cite{shikin}.  
This study was extended by us for the more general case where 
we consider the nonlinear term as an arbitrary function of all possible 
invariants generated from spinor bilinear forms. In that paper we also
studied the possibility of elimination of initial singularity especially for
the Kasner Universe~\cite{saha1}. For few years we studied 
the behavior of self-consistent NLSF in a B-I Universe~\cite{saha2,saha3} 
both in presence of perfect fluid and without it that was 
followed by the Refs.,~\cite{saha4,saha5,saha6} where we studied
the self-consistent system of interacting spinor and scalar fields.
The purpose of the paper is to study the role of the cosmological
constant ($\Lambda$) in the Lagrangian which together with 
Newton's gravitational constant ($G$) is considered as the fundamental
constants in Einstein's theory of gravity~\cite{einstein}. 

\section{Fundamental Equations and general solutions}

The Lagrangian for the self-consistent system of  spinor  and gravitational 
fields can be written as 
\begin{equation} 
L=\frac{R + 2 \Lambda}{2\kappa}+\frac{i}{2} 
\biggl[\bp \gamma^{\mu} \nabla_{\mu} \psi- \nabla_{\mu} \bar 
\psi \gamma^{\mu} \psi \biggr] - m\bp \psi + L_N,
\label{lag} 
\end{equation} 
with $R$- being the scalar curvature, $\kappa$- being  the  
Einstein's gravitational constant. The nonlinear term $L_N$
describes the self-interaction of a spinor field and in this particular 
case is chosen as an arbitrary function of $S= \bp \psi$, i.e.
$L_N = F(S)$. 

Variation of (\ref{lag}) with respect to spinor field gives
nonlinear Dirac equations
\begin{mathletters}
\label{speq}
\begin{eqnarray}
i\gamma^\mu \nabla_\mu \psi - \Phi \psi &=&\,0, \label{speq1} \\
 i \nabla_\mu \bp \gamma^\mu +  \Phi \bp &=&\,0, \label{speq2}
\end{eqnarray}
\end{mathletters}
with $\Phi = m - {\pr F}/{\pr S}$, whereas 
variation of (\ref{lag}) with respect to metric tensor
$g_{\mu\nu}$ gives the Einstein's field equation
\begin{equation}
R_{\nu}^{\mu} - \frac{1}{2}\,\delta_{\nu}^{\mu} R = - 8 \kappa 
T_{\nu}^{\mu} + \Lambda \delta_{\nu}^{\mu} 
\label{ee}
\end{equation} 
where $R_{\nu}^{\mu}$ is the Ricci tensor; $R = g^{\mu\,\nu} R_{\mu\,\nu}$
is the Ricci scalar; and $T_{\nu}^{\mu}$ is the energy-momentum tensor
of matter field defined as 
\begin{eqnarray}
T_{\mu}^{\rho}&=&\frac{i g^{\rho\nu}}{4}\bigl(\bp \gamma_\mu 
\nabla_\nu\psi + \bp\gamma_\nu\nabla_\mu\psi - \nabla_\mu\bp 
\gamma_\nu\psi - \nabla_\nu\bp\gamma_\mu\psi \bigr) \nonumber\\
& &-\delta_{\mu}^{\rho}L_{sp}.
\label{tem}
\end{eqnarray}
$L_{sp}$ with respect to (\ref{speq}) takes the form
\begin{equation}
L_{sp} = -\frac{1}{2}\Bigl(\bp \frac{\pr F}{\pr \bp} + \frac{\pr F}{\pr \psi}
\psi\Bigr) - F.
\label{lsp}
\end{equation}

In (\ref{speq}) and (\ref{tem}) $\nabla_\mu$ denotes the covariant 
derivative of spinor, having the form~\cite{zhelnorovich,brill} 
\begin{equation} 
\nabla_\mu \psi=\frac{\partial \psi}{\partial x^\mu} -\Gamma_\mu \psi, 
\label{cvd}
\end{equation} 
where $\Gamma_\mu(x)$ are spinor affine connection matrices. 
$\gamma$ matrices in the above equations are connected with 
the flat space-time Dirac matrices $\bg$ in the following way
$$ g_{\mu \nu} (x)= e_{\mu}^{a}(x) e_{\nu}^{b}(x) \eta_{ab}, 
\quad \gamma_\mu(x)= e_{\mu}^{a}(x) \bg_a,$$ 
where $\eta_{ab}= {\rm diag}(1,-1,-1,-1)$ and $e_{\mu}^{a}$ is a 
set of tetrad 4-vectors. 
 
Let us now choose Bianchi type-I space-time in the form~\cite{zeldovich} 
\begin{equation} 
ds^2 = dt^2 - a^2 dx^2 - b^2 dy^2 - c^2 dz^2, 
\label{BI}
\end{equation}
with $a,\,b$ and $c$ being the functions of $t$ only. 

For the space-time (\ref{BI}) Einstein equations (\ref{ee}) now read
\begin{mathletters}
\label{BID}
\begin{eqnarray}
\frac{\ddot b}{b} +\frac{\ddot c}{c} + \frac{\dot b}{b}\frac{\dot 
c}{c}&=&  8 \pi G T_{1}^{1} -\Lambda,\label{11}\\
\frac{\ddot c}{c} +\frac{\ddot a}{a} + \frac{\dot c}{c}\frac{\dot 
a}{a}&=&  8 \pi G T_{2}^{2} - \Lambda,\label{22}\\
\frac{\ddot a}{a} +\frac{\ddot b}{b} + \frac{\dot a}{a}\frac{\dot 
b}{b}&=&  8 \pi G T_{3}^{3} - \Lambda,\label{33}\\
\frac{\dot a}{a}\frac{\dot b}{b} +\frac{\dot b}{b}\frac{\dot c}{c} 
+\frac{\dot c}{c}\frac{\dot a}{a}&=&  8 \pi G T_{0}^{0} - \Lambda,
\label{00}
\end{eqnarray}
\end{mathletters}
where point means differentiation with respect to t. 
 
From
\begin{equation}
\Gamma_\mu (x)= 
\frac{1}{4}g_{\rho\sigma}(x)\biggl(\partial_\mu e_{\delta}^{b}e_{b}^{\rho} 
- \Gamma_{\mu\delta}^{\rho}\biggr)\gamma^\sigma\gamma^\delta, 
\label{gm}
\end{equation}
one finds
\begin{eqnarray} 
\Gamma_0&=&0, \quad 
\Gamma_1=\frac{1}{2}\dot a(t) \bg^1 \bg^0, \nonumber\\ \\
\Gamma_2&=&\frac{1}{2}\dot b(t) \bg^2 \bg^0,  
\Gamma_3=\frac{1}{2}\dot c(t) \bg^3 \bg^0. \nonumber
\end{eqnarray}
Flat space-time matrices $\bg$ we will choose in the form, 
given in~\cite{bogoliubov}. 

We will study the space-independent solutions to spinor field 
equation (7) so that $\psi=\psi(t).$ Setting
\begin{equation}
\tau = a b c = \sqrt{-g}
\label{taudef}
\end{equation}
in this case equation (\ref{speq1}) we write as
\begin{equation} i\bg^0 
\biggl(\frac{\partial}{\partial t} +\frac{\dot \tau}{2 \tau} \biggr) \psi 
- \Phi \psi = 0.
\label{spq}
\end{equation} 
Further putting  
$V_j(t) = \sqrt{\tau} \psi_j(t), \quad j=1,2,3,4,$ from (\ref{spq}) 
one deduces the following system of equations:  
\begin{mathletters}
\label{V}
\begin{eqnarray} 
\left. \begin{array}{c}
\dot{V}_{1} + i \Phi V_{1} = 0, \\
\dot{V}_{3} - i \Phi V_{3} = 0, 
\end{array}\right\}\\
\left. \begin{array}{c}
\dot{V}_{2} + i \Phi V_{2} = 0, \\
\dot{V}_{4} - i \Phi V_{4} = 0. 
\end{array}\right\}
\end{eqnarray} 
\end{mathletters}
From (\ref{V}) one easily obtains
\begin{equation}
\psi_{1,2} = \frac{C_{1,2}}{\sqrt{\tau}} e^{-i\Omega},\qquad
\psi_{3,4} = \frac{C_{3,4}}{\sqrt{\tau}} e^{i\Omega},
\label{psi}
\end{equation} 
where $\Omega = \int \Phi dt$ and $C_j$ are the constants of integration.

From (\ref{spq}) we will also find the equation for 
bilinear spinor form $S=\bp \psi$: 
\begin{equation} 
\dot S + \frac{\dot \tau}{\tau}S = 0,
\label{S} 
\end{equation}
with the solution
\begin{equation}
S = \frac{C_0}{\tau}
\label{st}
\end{equation}
where $C_0$ is the constant of integration. As one can see the constants
$C_0$ and $C_j$ are connected with each other as 
$$C_0 = C_{1}^{2} + C_{2}^{2} - C_{3}^{2} - C_{4}^{2}.$$

Putting (\ref{st}) into (\ref{tem}), we obtain the following expressions 
for the components of the energy-momentum tensor  
\begin{equation}
T_{0}^{0}=mS - F(S), \quad 
T_{1}^{1}=T_{2}^{2}=T_{3}^{3}= \frac{\pr F}{\pr S} S - F(S). 
\label{temc}
\end{equation}
From (\ref{temc}) in view of (\ref{st}) it is obvious that for the 
linear spinor field 
\begin{equation}
T_{0}^{0} = mS = mC_0/\tau, \quad 
T_{1}^{1} = T_{2}^{2} = T_{3}^{3} = 0. 
\label{temcl}
\end{equation}
The sign of $C_0$ is defined from the requirement of positivity of
energy density $T_{0}^{0}$ of linear spinor field. Hence, from (\ref{temcl})
emerges $C_0 > 0.$
In view of (\ref{temc}) from (\ref{BID}) we derive~\cite{saha1,saha2,saha3}
\begin{mathletters}
\label{abc}
\begin{eqnarray} 
a(t) &=& 
(D_{1}^{2}D_{3})^{1/3}\tau^{1/3}\, {\rm exp}[(2X_1 
+X_3)\,\vartheta], \label{a} \\
b(t) &=& 
(D_{1}^{-1}D_{3})^{1/3}\tau^{1/3}\, {\rm exp}[-(X_1 
-X_3)\,\vartheta], \label{b}\\
c(t) &=& 
(D_{1}D_{3}^{2})^{-1/3}\tau^{1/3}\, {\rm exp}[-(X_1 
+2X_3)\,\vartheta]. \label{c}
\end{eqnarray}
\end{mathletters}
where we define $\vartheta = \int\,[1/3 \tau (t)] dt$.

As one sees, both spinor field and metric functions are in some
functional dependent on $\tau$. Let us define the equation for $\tau$.
Summation of Einstein equations (\ref{11}), (\ref{22}),(\ref{33}) and 
(\ref{00}) multiplied by 3 gives
\begin{equation}
\ddot{\tau} = 
\frac{3\kappa \tau}{2} \Bigl(T_{1}^{1}+T_{0}^{0}\Bigr) - 3 \Lambda \tau 
\label{dtau}
\end{equation} 
The right-hand-side of (\ref{dtau}) is a function
of $\tau$ only, namely
$$\frac{3 \kappa}{2}\Bigl(m C_0 + \frac{\pr F}{\pr S} C_0 - 2 F \tau\Bigr)
- 3 \Lambda \tau := {\cal F} (\tau), $$
the solution to this equation is well-known for any arbitrary
function ${\cal F}(\tau)$~\cite{kamke} and can be written in quadrature
\begin{equation}
\int\,\frac{d \tau}{\sqrt{2\int {\cal F}(\tau) d \tau}} = t + t_0
\label{quadra}
\end{equation}
where $t_0$ is some constant that can be set zero. Given the explicit form of
the nonlinear term $F(S)$ from (\ref{quadra}) one finds the concrete 
solution for $\tau$. Thus the initial systems of
Einstein and Dirac equations have been completely integrated.

\section{Analysis of the results} 

In this section we shall analyze the general results obtained in the
previous section for concrete nonlinear term.

Let us consider the nonlinear term as a power function of $\tau$, precisely
$F(S) =\lambda S^n$ with  $\lambda$ being the coupling constant and $n>1$. 
Inserting $F(S)$ into (\ref{dtau})  one obtains
\begin{equation}
\ddot \tau = \frac{3\kappa C_0}{2} \biggl[m + \lambda (n-2) 
\frac{C_{0}^{n-1}}{\tau^{n-1}}\biggr] - 3 \Lambda \tau.  
\label{tn1}
\end{equation}
The first integral of (\ref{tn1}) has the form
\begin{equation}
\dot{\tau}^2 = 3 \kappa C_0 \Bigl[ m \tau - \lambda C_0^{n-1}/\tau^{n-2}
+ \tau_0 \Bigr] - 3 \Lambda \tau^2
\label{fi}
\end{equation}
with $\tau_0$ being some positive constant. Finally, we obtain 
\begin{equation}
\int \frac{\tau^{(n-2)/2} d \tau}{\sqrt{\kappa C_0 \tau_0 \tau^{n-2} +
\kappa C_0 m \tau^{n-1} - \Lambda \tau^n - \lambda C_0^n}} = \pm \sqrt{3}t.
\label{quadrature}
\end{equation}
Depending on the sign of $\Lambda$ and $\lambda$ we have the following
pictures. 

{\bf case 1.} $\Lambda = - \epsilon^2 < 0$, $\lambda > 0$. In this case for 
$n > 2$ and $t \to \infty$ we find
\begin{equation}
\tau (t) \approx e^{\sqrt{3} \epsilon t}
\label{c1}
\end{equation}  
Thus we see that the asymptotic behavior of $\tau$ does not depend on
$n$ and defined by $\Lambda$ - term. From (\ref{abc}) it is obvious that
the asymptotic isotropization takes place.

From (\ref{quadrature}) it also follows that $\tau$ cannot be zero
at any moment, since the intigrant turns out to be imaginary as $\tau$
approaches to zero. Thus the solution obtained is a nonsingular one
thanks to the nonlinear term in the Dirac equation and asymptotically
isotropic.

Let us go back to the energy density of spinor field. From 
\begin{equation}
T_{0}^{0} = \frac{mC_0}{\tau} - \frac{\lambda C_0^n}{\tau^n}
\end{equation} 
follows that at
\begin{equation}
\tau^{n-1} < \frac{\lambda C_0^{n-1}}{m}
\end{equation}
the energy density of spinor field becomes negative, which means 
that the absence of initial singularity in the considered cosmological 
solution appears to be consistent with the violation of the dominant 
energy condition in the Hawking-Penrose theorem~\cite{zeldovich}, 
since in this case
\begin{equation}
T_{1}^{1} = T_{2}^{2} = T_{3}^{3} = \frac{\lambda (n-1) C_0^n}{\tau^n} > 0.
\end{equation}
Consider the linear case with $\lambda = 0$. Then from (\ref{quadrature})
follows
\begin{equation}
\tau (t) = \frac{1}{4} e^{\sqrt{3}\epsilon t} + e^{-\sqrt{3}\epsilon t}
\Bigl(\frac{\kappa^2 m^2 C_{0}^{2}}{4 \epsilon^4} -
\frac{\kappa C_0 \tau_0}{\epsilon^2}\Bigr) - \frac{\kappa m C_0}{2 \epsilon^2}.
\label{tl0}
\end{equation}
As one sees
\begin{equation}
\lim\limits_{t \to \infty} \tau \approx \frac{1}{4} e^{\sqrt{3}\epsilon t},
\end{equation}
that coincides with (\ref{c1}). from (\ref{tl0}) follows
\begin{equation}
\tau (0) = \frac{1}{4}\Bigl(1 - \frac{\kappa m C_0}{\epsilon^2}\Bigr)^2
- \frac{\kappa C_0 \tau_0}{\epsilon^2},
\label{tau0}
\end{equation}
that means, $\tau (0)$ is defined by the relation between the constants.

{\bf case 2.} $\Lambda > 0$ and $\lambda > 0$. For $n > 2$ 
(\ref{quadrature}) admits only nonsingular oscillating solutions, since
$\tau > 0$ and bound from above. Consider the case with $n = 4$ and
for simplicity set $m = 0$. Then from (\ref{quadrature}) one gets
\begin{equation}
\tau(t) = \frac{1}{\sqrt{2 \Lambda}} \Bigl[\kappa C_0 \tau_0
+ \sqrt{\kappa^2 C_0^2 \tau_0^2 + 4 \Lambda \lambda C_0^4}\, 
{\rm sin} 2 \sqrt{3 \Lambda} t \Bigr]^{1/2}.
\end{equation}
For a spinor field with $\Lambda > 0$ and $\lambda > 0$
and $n = 10$ a perspective view of $\tau$ is shown in FIG. 1.
Period for massive field ($m\ne 0$) is greater than that for 
massless one ($m=0$). The initial value of $\tau$ has been taken to be
unit, i.e., $\tau (0) = 1$. For $\tau (0) = 10^{-2}$ for example, $\tau$
increases tremendously at initial steps (from $10^{-2}$ to $10^{13}$ in 
first step) and amplitude in that case is $10^{16}$, whereas the value of
$n$ (order of nonlinearity) defines the period (the more is $n$ the less
is period). 

\vspace*{.5cm}
\begin{figure}
\hspace{.1cm}\epsfig{file=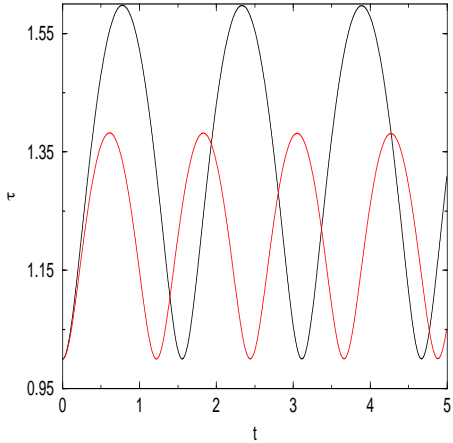,height=6cm,width=6cm,angle=0}
\vspace{.5cm}
\caption{Perspective view of $\tau$ showing the initially
nonsingular and oscillating behavior of the solutions}
\end{figure}

{\bf case 3.} $\Lambda < 0$ and $\lambda < 0$. The solution is
singular at initial moment, that is
\begin{equation}
\lim\limits_{t \to 0} \tau \approx [\sqrt{-3 \lambda n^2 C_0^n/4}t]^{2/n}
\label{c3}
\end{equation} 
and at $t \to \infty$ asymptotic isotropization takes place since
\begin{equation}
\lim\limits_{t \to \infty} \tau \approx e^{\sqrt{3 \Lambda} t}.
\end{equation} 

{\bf case 4.} $\Lambda > 0$ and $\lambda < 0$. Solution is initially
singular and coincides with (\ref{c3}) and bound from the above, i.e., 
oscillating, since
\begin{equation}
\lim\limits_{t \to \infty} \tau \approx {\rm sin} \sqrt{3 \Lambda} t.
\end{equation} 

\section{Conclusion}
Within the framework of the simplest nonlinear model of spinor field 
it has been shown that the $\Lambda$ term plays very important role 
in Bianchi-I cosmology. In particular, it invokes oscillations in the
model which is not the case when $\Lambda$ term remain absent. 
Growing interest in studying the role $\Lambda$ term by present day 
physicists of various discipline witnesses its fundamental value.
For details on time depending $\Lambda$ term one may consult~\cite{sahal}
and references therein.

\end{document}